\documentclass[11pt]{amsart}
\usepackage{color}
\usepackage{amsfonts,amsmath,amsthm,amssymb,latexsym}
\usepackage{tikz}
\usepackage{cancel}
\usepackage{graphicx}
\usepackage{enumitem}
\usepackage{geometry}

\usepackage{algorithm}
\usepackage{algorithmic}

\usepackage{mathrsfs}

\usetikzlibrary{matrix,calc,shapes}
\tikzset{
	treenode/.style = {shape=rectangle, rounded corners,
		draw, anchor=center, align=center,
		top color=white, bottom color=blue!20,
		inner sep=1ex},
	decision/.style = {treenode, diamond, inner sep=0pt},
	root/.style     = {treenode, bottom color=red!30},
	env/.style      = {treenode},
	finish/.style   = {root, bottom color=green!40},
	dummy/.style    = {circle,draw}
}

\DeclareMathOperator{\ord}{ord}

\DeclareMathOperator{\prem}{prem}
\DeclareMathOperator{\Sol}{Sol}
\DeclareMathOperator{\Param}{IFP}
\DeclareMathOperator{\Pla}{Places}

\newcommand{\im}{\mathrm{Im}}
\newcommand{\K}{\mathbb{K}}
\newcommand{\LL}{\mathbb{L}}
\newcommand{\KK}{\overline{\K}}
\newcommand{\Q}{\mathbb{Q}}
\newcommand{\Z}{\mathbb{Z}}
\newcommand{\cc}{\mathscr{C}}
\newcommand{\qq}{\mathbf{p}_0}
\newcommand{\para}{\vspace*{2mm}}

\newtheorem{theorem}{Theorem}[section]

\newtheorem{lemma}[theorem]{Lemma}

\theoremstyle{definition}

\theoremstyle{remark}

\numberwithin{equation}{section}
\theoremstyle{remark}

\begin{document}

\date{\today}
\title[First Order Autonomous AODEs]{Puiseux Series and Algebraic Solutions of First Order Autonomous AODEs -- A MAPLE Package}

\author[F. Boulier]{Fran{\c{c}}ois Boulier}
\address{Univ. Lille, CNRS, Centrale Lille, Inria, UMR 9189 - CRIStAL, Lille, France.}
\email{francois.boulier@univ-lille.fr}

\author[J. Cano]{Jos\'{e} Cano}
\address{Dpto. Algebra, an\'alisis matem\'atico, geometr\'{\i}a y topolog\'{\i}a, Universidad de Valladolid, Spain.}
\email{jcano@agt.uva.es}

\author[S. Falkensteiner]{Sebastian Falkensteiner}
\address{Research Institute for Symbolic Computation (RISC), Johannes Kepler University Linz, Austria.}
\email{falkensteiner@risc.jku.at}

\author[J.R. Sendra]{J.Rafael Sendra}
\address{Dpto. de F\'{\i}sica y Matem\'aticas, Universidad de Alcal\'a, Madrid, Spain.}
\email{rafael.sendra@uah.es}

\thanks{The second author was partially supported by MTM2016-77642-C2-1-P (AEI/FEDER, UE). 
	The third and fourth authors were partially supported by FEDER/Ministerio de Ciencia, Innovaci\'{o}n y Universidades Agencia Estatal de Investigaci\'{o}n/MTM2017-88796-P (Symbolic Computation: new challenges in Algebra and Geometry together with its applications).
	The third author was also supported by the Austrian Science Fund (FWF): P 31327-N32.
	The fourth author is member of the Research Group ASYNACS (Ref. CT-CE2019/683).}

\begin{abstract}
There exist several methods for computing exact solutions of algebraic differential equations. 
Most of the methods, however, do not ensure existence and uniqueness of the solutions and might fail after several steps, or are restricted to linear equations. 
The authors have presented in previous works a method to overcome this problem for autonomous first order algebraic ordinary differential equations and formal Puiseux series solutions and algebraic solutions. 
In the first case, all solutions can uniquely be represented by a sufficiently large truncation and in the latter case by its minimal polynomial.

The main contribution of this paper is the implementation, in a \texttt{MAPLE} package named \texttt{FirstOrderSolve}, of the algorithmic ideas presented therein. 
More precisely, all formal Puiseux series and algebraic solutions, including the generic and singular solutions, are computed and described uniquely. 
The computation strategy is to reduce the given differential equation to a simpler one by using local parametrizations and the already known degree bounds.
\end{abstract}

\maketitle

\noindent \textbf{keywords}
Maple, Symbolic computation, Algebraic differential equation, Formal Puiseux series solution, Algebraic solution.

\section{Introduction}
The problem of finding power series solutions of ordinary differential equations has been extensively studied in the literature.
A method to compute generalized formal power series solutions, i.e. power series with real exponents, and to describe their properties is the Newton polygon method.
A description of this method is given in~\cite{fine1889,fine1890} and more recently in~\cite{GrSi:1991,DoraJung1997,Aroca2000}.
In~\cite{Cano2005}, the second author, using the Newton polygon method, gives a theoretical description of all generalized formal power series solutions of a non-autonomous first order ordinary differential equation as a finite set of one parameter families of generalized formal power series.
This description of the solutions is in general not algorithmic by several reasons.
One of them is that there is no bound on the number of terms which have to be computed in order to guarantee the existence of a generalized formal power series solution when extending a given truncation of a determined potential solution.
Also the uniqueness of the extension can not be ensured a-priori.

In~\cite{cano2019existence} this problem has been overcome by the authors for autonomous first order differential equations by using a local version of the algebro-geometric approach introduced in~\cite{feng2004rational}.

In~\cite{RISC5589} they derive an associated differential system to find rational general solutions of non-autonomous first order differential equations by considering rational parametrizations of the implicitly defined curve.
We instead consider its places and obtain an associated differential equation of first order and first degree which can be transformed into an equation of a very specific type~\cite{BriotBouquet-Reserches}.
Using the known bounds for computing places of algebraic curves (see e.g.~\cite{Duval1989}), existence and uniqueness of the solutions and the termination of our computations can be ensured.

In~\cite{aroca2005algebraic} the results of~\cite{feng2004rational,feng2006polynomial} are generalized to algebraic solutions. 
It is well known that algebraic solutions can be represented as Puiseux series. 
The advantage is that they can be fully described by its minimal polynomial. 
In this package we mainly follow~\cite{aroca2005algebraic}, but we use an adapted version of the algorithm there for deciding the existence of algebraic solutions and computing all of them in the affirmative case.

\section{Theoretical and Algorithmic Framework}
In this section we recall the main notions and results that are used in our implementations. 
For further details we refer to~\cite{cano2019existence} in the case of formal Puiseux series and to~\cite{aroca2005algebraic} in the case of algebraic solutions.

\para

Let $\K$ be a computable field of characteristic zero such as the rational numbers $\Q$ and let us denote by $\KK$ its algebraic closure. 
Let us consider the differential equation
\begin{equation}\label{eq-main}
F(y,y')=0,
\end{equation}
where $F \in \K[y,p]$ is square-free and non-constant in the variables $y$ and $p$. 
We are looking for formal Puiseux series and algebraic solutions of~\eqref{eq-main}. 
In the case of formal Puiseux series solutions we will represent the full series by a sufficiently large truncation such that existence and uniqueness are guaranteed. 
In the case of algebraic solutions we look for its minimal polynomial.

We associate to~\eqref{eq-main} the affine algebraic curve $C(F) \subset \KK^2$ defined by the zero set of $F(y,p)$ in $\KK^2$.
We denote by $\cc(F)$ the Zariski closure of $C(F)$ in $\KK_{\infty}^2$, where $\KK_{\infty}=\KK\cup \{\infty\}$ denotes the one-point compactification of $\KK$. 
In the case of formal Puiseux series solutions we will look for local parametrizations of $\cc(F)$ and in the case of algebraic solutions for algebraic parametrizations, respectively.

\subsection{Formal Puiseux Series Solutions}
Formal Puiseux series can either be expanded around a finite point or at infinity.
In the first case, since equation~\eqref{eq-main} is invariant under translation of the independent variable, without loss of generality we can assume that the formal Puiseux series is expanded around zero and it is of the form $\varphi(x)=\sum_{j\geq j_0}a_j\,x^{j/n}$, where $a_j\in \KK,$ $n \in \Z_{>0}$ and $j_0\in \Z$.
In the case of infinity we can use the transformation $x=1/z$ obtaining the (non-autonomous) differential equation $F(y(z),-z^2y'(z))=0$.
In order to deal with both cases in a unified way, we will study equations of the type
\begin{equation}\label{eq-infinity}
F(y(x),(1-h)x^hy'(x))=0,
\end{equation}
with $h \in \{0,2\}$ and its formal Puiseux series solutions expanded around zero.
We note that for $h=0$ equation~\eqref{eq-infinity} is equal to~\eqref{eq-main} and for $h=2$ the case of formal Puiseux series solutions expanded at infinity is treated.

We use the notations $\LL[[x]]$ for the ring of formal power series, $\LL((x))$ for its fraction field and $\LL((x))^*=\bigcup_{n \geq 1} \LL((x^{1/n}))$ for the field of formal Puiseux series expanded at zero with coefficients in some field $\LL$. 
We call the minimal natural number $n$ such that $\varphi(x)$ belongs to $\LL((x^{1/n}))$ the \textit{ramification order} of $\varphi(x)$.
Moreover, for $\varphi(x)=\sum_{j\geq j_0}a_j\,x^{j/n}$ with $a_{j_0} \neq 0$ we call $j_0/n \in \Q$ the order of $\varphi$, denoted by $\ord_x(\varphi(x))$, and set $\ord_x(\varphi(x))=\infty$ for $\varphi=0$.

Additionally to~\eqref{eq-infinity} we may require that a formal Puiseux series solution $y(x)$ of~\eqref{eq-infinity} fulfills the initial conditions $y(0)=y_0,((1-h)x^hy'(x))(0)=p_0$ for some fixed $\qq=(y_0,p_0) \in \KK_{\infty}^2$. 
In the case where $y(0)=\infty$, $\tilde{y}(x)=1/y(x)$ is a Puiseux series solution of a new first order differential equation of the same type, namely the equation given by the numerator of the rational function $F(1/y,-(1-h)x^hp/y^2)$, and $\tilde{y}(0)\in \KK$.
Therefore, in the sequel, we may assume that $\qq\in \KK\times \KK_{\infty}$.

\subsubsection*{Formal Parametrizations.}
Let us recall some classical terminology on local parametrizations of algebraic curves and its algorithmic aspects, for further details see e.g.~\cite{walker1950algebraic,Duval1989}.

\para

A \textit{formal parametrization} centered at $\qq\in \cc(F)$ is a pair of formal Puiseux series $A(t)\in \KK((t))^2\setminus \KK^2$ such that $A(0)=\qq$ and $F(A(t))=0$.
In the set of all formal parametrizations of $\cc(F)$ we introduce the equivalence relation $\sim$ by defining $A(t) \sim B(t)$ if and only if there exists a formal power series $s(t)\in \KK[[t]]$ of order one such that $A(s(t))=B(t)$.
A formal parametrization is said to be \textit{irreducible} if it is not equivalent to another one in $\KK((t^m))^2$ for some $m>1$.
An equivalence class of an irreducible formal parametrization $(a(t),b(t))$ is called a \textit{place} of $\cc(F)$ centered at the common center point $\qq$ and is denoted by $[(a(t),b(t))]$.

\para

Let $\Param(\qq)$ denote the set of all irreducible formal parametrizations of $\cc(F)$ at $\qq$ and $\Pla(\qq)$ containing the places of $\cc(F)$ centered at $\qq$.
Computationally we have to truncate the formal parametrizations. 
There are bounds presented in~\cite{Duval1989,RISC4119} such that
\begin{enumerate}
	\item the truncations of the formal parametrizations $(a(t),b(t))$ at $\qq$ are in one-to-one correspondence to $\Pla(\qq)$;
	\item the orders $\ord_t(a(t)-y_0),\ord_t(b(t))$ are determined;
	\item no further extension of the ground field for computing the following coefficients have to be done.
\end{enumerate}
For the monic case, a bound for the number of steps to compute it is $2(\deg_p(F)-1) \deg_y(F)+1$ or the Milnor number. 
For the general case, a similar bound based on the degree of the polynomial can be derived. 

Let us note that the solutions of~\eqref{eq-infinity} will be independent of the chosen representative of a place. 
Hence, regarding uniqueness of the prolongation, number of field extensions, etc. it does not matter which local parametrization we chose (for example classical Puiseux parametrizations or rational Puiseux parametrization~\cite{Duval1989}). 
For representing the solution parametrizations, which is not the goal of the current paper, however, it would be relevant.

\subsubsection*{Puiseux Solution Place.}
Let $\Sol_{\KK((x))^*}(\qq)$ be the set containing the non-constant formal Puiseux series solutions of equation~\eqref{eq-infinity}, expanded at zero, with coefficients in $\KK$ and with $\qq$ as initial values. 
Then the mapping $\Delta:\Sol_{\KK((x))^*}(\qq) \longrightarrow \Param(\qq)$ defined as $$\Delta(y(x))=\left(y(t^{n}),(1-h)t^{hn}\,\frac{d\, y}{d\, x}(t^{n})\right),$$ where $n$ is the ramification order of $y(x)$, is well-defined and injective. 
Moreover, we denote by $\delta: \Sol_{\KK((x))^*}(\qq) \longrightarrow \Pla(\qq)$ the map $\delta(y(x))=[\Delta(y(x))]$.

An irreducible formal parametrization $A(t)\in\Param(\qq)$ is called a \textit{solution parametrization} of~\eqref{eq-main} if $A$ is in the image $\im(\Delta)$.
Similarly, a place in $\im(\delta)$ is called a \textit{(Puiseux) solution place}.

It can be shown that for solution parametrizations $(a(t),b(t))\in\Param(\qq)$, corresponding to a solution with ramification index $n$, it holds that
\begin{align}\label{eq:nec_eq_order_ramification}
n(1-h)=\ord_t(a(t)-y_0)-\ord_{t}(b(t)).
\end{align}
This condition is invariant for the representative of a place. 
In particular, all Puiseux series solutions in the same solution place have the same ramification order. 
It turns out that condition~\eqref{eq:nec_eq_order_ramification} is already sufficient for solution places at $\qq$ with $y_0 \in \KK$.
Let us highlight this statement (see Theorem 10 in~\cite{cano2019existence}):
\begin{theorem}\label{theorem:characterizationSolPlace}
	Let $\mathcal{P}=[(a(t),b(t))]\in \Pla(\qq)$ and $h=0$.
	Then $\mathcal{P}$ is a solution place if and only if equation~\eqref{eq:nec_eq_order_ramification} holds for an $n \in \Z_{>0}$.
	In the affirmative case the ramification order of $\mathcal{P}$ is equal to $n$.
\end{theorem}

Also the solutions with $h=2$ can be computed algorithmically. 
For this purpose let us give in the following some insight into to proof of Theorem~\ref{theorem:characterizationSolPlace}.

Let $\LL$ be a subfield of $\KK$. 
For a given parametrization $(a(t),b(t))) \in \LL((t))^2$ satisfying \eqref{eq:nec_eq_order_ramification}, our strategy is to find $s(t) \in \KK[[t]]$ with $\ord_t(s(t))=1$ such that $(a(s(t)),b(s(t)))$ satisfies the \textit{associated differential equation}
\begin{equation}\label{eq:DE_2}
a'(s(t))\cdot s'(t)=n(1-h)\,t^{n(1-h)-1}\,b(s(t)).
\end{equation}
Let $k=\ord_t(a(t)-y_0), r=\ord_t(b(t))$ and $n(1-h)=k-r>0$. 
By transforming~\eqref{eq:DE_2} into an equation of Briot-Bouquet type~\cite{BriotBouquet-Reserches}, the solutions $s(t)=\sum_{i=1}^{\infty} \sigma_i\,t^i$ fulfill the following items.
\begin{enumerate}
	\item If $h=0$, there are exactly $n$ solutions where $\sigma_1^{n} \in \LL$ and $\sigma_i \in \LL(\sigma_1)$ are uniquely determined for $i>1$.
	\item If $h=2$, there is no solution or up to $n$ one-parameter families of solutions with $\sigma_1^{n} \in \LL$, $\sigma_{r-k} \in \LL$ is a free parameter; $\sigma_2,\ldots,\sigma_{r-k-1} \in \LL(\sigma_1)$ and for $i>r-k$ the coefficients $\sigma_{i} \in \LL(\sigma_1,\sigma_{r-k})$ are uniquely determined.
\end{enumerate}
After computing the solutions $s(t)$ of the associated differential equation, we obtain the solutions of the original differential equation by $a(s(x^{1/n}))$.

\subsubsection*{Solution Truncations.}
Since we cannot compute all coefficients of the Puiseux series solution, we have to truncate at some point.
A point $\qq=(y_0,p_0) \in \cc(F)$ is called a \textit{critical curve point} if $p_0 \in \{0,\infty\}$ or $\frac{\partial F}{\partial p}(\qq)=0$ or $y_0=\infty$.
Under our assumptions, the set of critical curve points is finite. 
The only formal Puiseux series solution with non-critical $\qq$ as initial tuple is a formal power series and its determined solution truncation is given by $y_0+p_0x$.

Assume that $\qq \in \cc(F)$ is a critical curve point. 
Then, by the properties of the solutions of the associated differential equations, the bound on the number of computational steps from~\cite{Duval1989,RISC4119} also holds for the computation of the determined solution truncations.
In particular, equation~\eqref{eq:nec_eq_order_ramification} can be checked, no further extensions of the ground field for computing the coefficients are necessary and the ramification index is determined.
For the general case, a similar bound based on the degree of the polynomial can be derived and we will write $N$ for such a bound.


\begin{algorithm}[H]
	\label{alg-solutions puiseux}
	\caption{PuiseuxSolve}
	\begin{algorithmic}[1]
		\REQUIRE A first-order AODE $F(y,y')=0$, where $F\in \K[y,p]$ is square-free with no factor in $\K[y]$ or $\K[p]$.
		\ENSURE A set consisting of all solution truncations of $F(y,y')=0$ (expanded around zero and around infinity).
		\STATE If $(\infty,\infty) \in \cc(F)$, then perform the transformation $\tilde{y}=1/y$ and apply the following steps additionally to the numerator of $F(1/y,-p/y^2)$ and $\qq=(0,0)$ in order to obtain the solutions of negative order.
		\STATE Compute the set of critical curve points $\mathcal{B}(F)$ (for $y_0 \in \KK$) and $\mathbb{V}(F(y,0))$ (for $y_0=\infty$).
		\STATE For every point $(y_0,p_0) \in \cc(F) \setminus \mathcal{B}(F), y_0 \neq \infty$, a determined solution truncation is $y_0+p_0x$.
		\STATE Add to the output the constant solutions $y(x)=y_0$ corresponding to $(y_0,0) \in \cc(F), y_0 \neq \infty$.
		\STATE For every place centered at a critical curve point $\qq=(y_0,p_0) \in \mathcal{B}(F)$ and $(\infty,p_0) \in \cc(F)$, compute the first $N$ terms of a formal parametrization $(a(t),b(t))$.
		\STATE Check equation~\eqref{eq:nec_eq_order_ramification}: In the negative case, $[(a(t),b(t)]$ is not a solution place.
		\STATE In the affirmative case, compute the first $N$ terms of the solutions $s(t)$ of \eqref{eq:DE_2}.\\
		For $y_0 \neq \infty$ there exists exactly $n$ solutions. For $y_0=\infty$ the associated differential equation is either unsolvable or contains a free parameter.
		\STATE The first $N$ terms of $a(s(x^{1/n}))$ are the solution truncations with $\qq$ as initial tuple.
	\end{algorithmic}
\end{algorithm}

For finite initial values we are able to ensure uniqueness of the extension of the truncated Puiseux series solutions (see also~\cite{cano2019existence}[Theorem 14]). 
In the case of $y_0=\infty$ some truncations may coincide for specific values obtained in the solution of the reparametrization.

\subsection{Algebraic Solutions}
In this section we consider a subclass of formal Puiseux series, namely algebraic series. 
These are $y(x) \in \KK((x))^*$ such that there exists a non-zero $G \in \KK[x,y]$ with $G(x,y(x))=0$. 
Since the field of formal Puiseux series is algebraically closed, all algebraic solutions can be represented as (formal) Puiseux series.

In~\cite{aroca2005algebraic} a bound on the degree of algebraic general solutions is given. 
There the authors indicate how to use these results in order to compute all algebraic solutions of such a given differential equation. 
A more detailed proof of this fact can be found in~\cite{FalkensteinerThesis}.

\para

The first important observation is that if there exists one non-constant algebraic solution of~\eqref{eq-main}, then all of them can be found easily by a shift in the minimal polynomial (see \cite{FalkensteinerThesis}[Theorem 4.1.22]).
\begin{theorem} \label{THM:AllAlgebraic}
	Let $F \in \K[y,y']$ be irreducible and let $y(x)$ be a non-constant solution of $F=0$ algebraic over $\KK(x)$ with minimal polynomial $G \in \KK[x,y]$. 
	Then all formal Puiseux series solutions $\Sol_{\KK((x))^*}(F)$ are algebraic and given by $G(x+c,y)$, where $c \in \KK$.
\end{theorem}

The second important computational aspect is the degree bound on the solutions~\cite{aroca2005algebraic}[Theorem 3.4, Theorem 3.8]:
\begin{theorem} \label{THM:DegreeBound}
	Let $F \in \K[y,y']$ be irreducible and let $y(x)$ be a non-constant solution of $F=0$ algebraic over $\KK(x)$ with minimal polynomial $G \in \KK[x,y]$. 
	Then $$\deg_x(G)=\deg_p(F), \quad \deg_y(G) \leq \deg_y(F)+\deg_p(F).$$
\end{theorem}

The third result is used to construct candidates of algebraic
solutions:
\begin{lemma}\label{lemma:ReconstructingABounds}
	Let $G(x,y)\in \mathbb{K}[x,y]$ be an irreducible polynomial with $d_x=\deg_xG, d_y=\deg_y G$. 
	Let	$y(x)$ be a Puiseux series solution of $G(x,y)=0$ expanded at $x=0$ with $\ord_x(y(x))=\nu$. 
	Let $\nu'=\min\{\nu,0\}$ and write $y(x)=\bar{y}(x)+\varphi(x)$ with $\ord_x(\varphi(x))>N>0$ where
		\begin{align}
		N &\geq 2\,d_x\,d_y-2\,\nu'\,(d_y-1).\label{eq:ReconstructingSolBoundN}
		\end{align}
	Assume that $A(x,y)\in \mathbb{K}[x,y], \deg_xA\leq d_x$, $\deg_y A\leq d_y$ is of minimal degree such that
	\begin{align}\label{eq:ReconstructingSolBoundA}
	\ord_x(A(x,\bar{y}(x)) &> 2\,d_x\,d_y-\nu'(d_y-1),
	\end{align}
	holds. 
	Then $A(x,y)$ is, up to a constant factor, equal to $G(x,y)$.
\end{lemma}
\begin{proof}
	Let $R(x)$ be the resultant of $G(x,y)$ and $A(x,y)$ with respect to $y$. 
	It is well known that there exist polynomials $B(x,y)$, $C(x,y)$ with $\deg_yB<d_y$, $\deg_yC<d_y$ such that
	$$G(x,y)\,B(x,y)+A(x,y)\,C(x,y)=R(x).$$
	Evaluating at $\bar{y}(x)$ we obtain
	\begin{equation}\label{eq:resultant}
	G(x,\bar{y}(x))\,B(x,\bar{y}(x))+A(x,\bar{y}(x))\,C(x,\bar{y}(x))=R(x). 
	\end{equation}
	Since $\nu' \le 0$, it follows that $\ord_x C(x,\bar{y}(x))\geq \nu'\,\deg_yC\geq \nu'(d_y-1)$ and similarly for $B(x,\bar{y}(x))$. 
	Hence, by \eqref{eq:ReconstructingSolBoundA}, we have that $$\ord_x(A(x,\bar{y}(x))\,C(x,\bar{y}(x)))> 2\,d_x\,d_y.$$ 
	Let us proof that $\ord_x(G(x,\bar{y}(x))>\nu'(d_y-1)+N$.
	Taking the Taylor series of $G(x,\bar{y}(x)+\varphi(x))$ and because $G(x,\bar{y}(x)+\varphi(x))=0$, we have:
	\begin{align*}
	G(x,\bar{y}(x))=-\sum_{j=1}^{d_y}\frac{1}{j!}\frac{\partial^{j} G}{\partial y^{j}}(x,\bar{y}(x))\,\varphi(x)^{j}. 
	\end{align*}
	The order in $x$ of each term on the right hand side of above equation is greater than $\nu'\,(d_y-1)+N$, so it is for the left
	hand side. 
	Now, because of \eqref{eq:ReconstructingSolBoundN}, we have that
	\begin{align*}
	\ord_x(G(x,\bar{y}(x))\,B(x,\bar{y}(x)))> \nu'(d_y-1)+N+\nu'(d_y-1)\geq 2\,d_x\,d_y.
	\end{align*}
	Hence, the left hand side of \eqref{eq:resultant} has order greater than $2\,d_x\,d_y$ and the right hand side is a polynomial of degree less than or equal to $2\,d_x\,d_y$. 
	Hence, $R(x)=0$, and therefore, $G(x,y)$ and $A(x,y)$ have a common factor. 
	Since $G(x,y)$ is an irreducible polynomial, it is a factor of $A(x,y)$. 
	Then, by the degree conditions on $A(x,y)$, the statement follows.
\end{proof}

In \cite{aroca2005algebraic} the method of detecting candidates $G(x,y)$ for
algebraic solutions of the differential equations $F(y,y')=0$ consists 
by computing $\bar{y}(x)$, the first $N$ terms of a power series solution $y(x)$ of the
differential equations $F(y,y')=0$, with a regular curve point of $\cc(F)$ as initial tuple. 
Hence, in this case the solution $y(x)$ is of order $0$ and $\nu'=0$. 
Choose $N>d_x\,d_y$ and construct, by solving a linear system of equations, a polynomial $A$ fulfilling the properties~\eqref{eq:ReconstructingSolBoundN} and~\eqref{eq:ReconstructingSolBoundA}. 
This approach reduced the number of formal power series solutions that we
can use to construct a candidate. 
Lemma~\ref{lemma:ReconstructingABounds}
allows to choose   any Puiseux series solutions of the differential equations
and reduce the computational cost.

Once  a candidate $A(x,y)$ is
detected, we can check whether it is an actual algebraic solution of the
differential equation $F(y,y')=0$ by checking whether the differential pseudo
remainder of $F(y,y')$ with respect $A(x,y)$ is zero.
These results lead to the following algorithm.

\begin{algorithm}[H]
	\caption{AlgebraicSolve}
	\label{alg-solutions algebraic}
	\begin{algorithmic}[1]
		\REQUIRE A first-order AODE $F(y,y')=0$, where $F\in \K[y,p]$ is irreducible over $\KK(y)$.
		\ENSURE The minimal polynomial of an algebraic solution of
		$F(y,y')=0$, describing all solutions, if it exists.
		
		\STATE Compute the minimal number of terms of all Puiseux
		solutions of $F(y,y')=0$ using PuiseuxSolve and choose one of
		them, denote it by $\hat{y}(x)$. Let $\nu$ be its order,
		$\nu'=\min(\nu,0)$ and $n$ its
		ramification index.
		\STATE Let $d_x=\deg_{p}F$ and $d_y=\deg_yF+\deg_pF$.
		\STATE Compute the prolongation $\bar{y}(x)$ of $\hat{y}(x)$ up to order
		$N=2\,d_x\,d_y-2\,\nu'\,(d_y-1)+1/n$.
		\STATE Compute $A(x,y) \in \KK[x,y]$ fulfilling the required
		conditions from Lemma~\ref{lemma:ReconstructingABounds}
		by an ansatz of unknown coefficients and solving the resulting linear system.
		\STATE Check whether $\prem(F,A)=0$. If so, then $A(x,y)$ is an
		actual solution. Otherwise there exists no algebraic solution. 
	\end{algorithmic}
\end{algorithm}

\section{The Package FirstOrderSolve}
In this section, we present the structure and content of the \texttt{MAPLE} package \texttt{FirstOrderSolve}. 
It consists several procedures that implement in particular the algorithms PuiseuxSolve and AlgebraicSolve described above. 
This package computes the Puiseux series solutions and algebraic solutions of first order autonomous AODEs with coefficients in an algebraic extension field of $\Q$.

\subsection{Overview of the Software Structure}
The created \texttt{MAPLE} package is initialized by the command

\texttt{> with(FirstOrderSolve):}\\
The main procedures are
\begin{itemize}
	\item \texttt{SolutionTruncations}: for computing all formal Puiseux series solutions (Algorithm PuiseuxSolve);
	\item \texttt{AlgebraicSolution}: for computing the minimal polynomial of the algebraic solutions (Algorithm AlgebraicSolve);
	\item \texttt{GenericSolutionTruncation}: for computing a truncation of the solutions with non-critical initial tuple;
	\item \texttt{ProlongSolutionTruncation}: for prolonging the solution truncations up to a higher degree.
\end{itemize}
These four commands are public to the user. 
The package is divided into several sub-packages \texttt{BriotBouquetSolve, LocalSolve, AlgebraicSolve}, which are not accessible for the user, and uses the hierarchy scetched below.

\begin{figure}[H]
	\centering
	\begin{tikzpicture}[-latex]
	\matrix (chart)
	[matrix of nodes, column sep = 3em, row sep = 5ex, column 1/.style = {nodes={decision}}, column 2/.style = {nodes={env}}]
	{
		& |[root]| FirstOrderSolve & \\
		|[treenode]| AlgebraicSolve & & |[treenode]| LocalSolve \\
		& |[finish]| BriotBouquetSolve &  \\
	};
	\draw
	(chart-1-2) edge (chart-2-1)
	(chart-1-2) edge (chart-2-3)
	(chart-2-1) edge (chart-3-2)
	(chart-2-1) edge (chart-2-3)
	(chart-2-3) edge (chart-3-2);
	\end{tikzpicture}
	\caption{The hierarchy of the package.}
\end{figure}
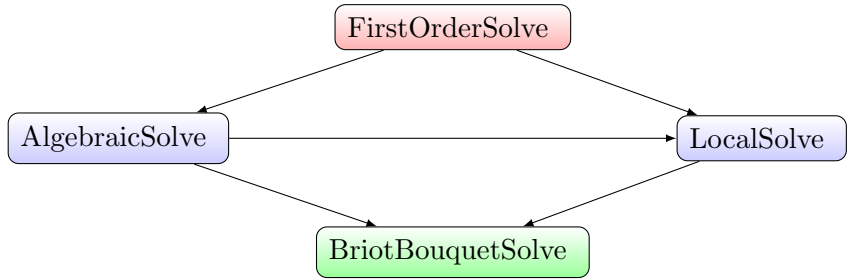

The main commands in the sub-packages are the following.
\begin{itemize}
	\item \texttt{ParametrizationSetAlgCurve}: for computing the formal parametrizations of an implicitly defined algebraic curve by using the command \texttt{algcurves:-puiseux};
	\item \texttt{ReparametrizationSet}: for computing the solutions of the associated differential equation by using \texttt{BriotBouquetSolve};
	\item \texttt{BriotBouquetSolve}: for computing the unique solution of a first-order differential equation in quasi-solved form (which is called an equation of Briot-Bouquet type~\cite{BriotBouquet-Reserches}[Section 80,86]); this procedure is using a Newton type algorithm for solving the resulting linear system in several variables;
\end{itemize}

In the following, we give a description of the procedures in the package \texttt{FirstOrderSolve}. 
The package is available at \texttt{risc.jku.at/sw/firstordersolve/}, where a more detailed information on the commands is presented in the help.

\subsection{Description of the Software Components}

{\bf \texttt{> SolutionTruncations}}\\
Computes all Puiseux series solutions of a given first order autonomous ordinary differential equation.\\
Since the equation is autonomous, the translation of the independent variable by any constant in a solution is again a solution. 
Hence, the only relevant expansion points are $0$ and infinity.\\
The solutions expanded at $0$ can be split into two sets: a generic solution and a set of particular solutions. 
The generic solution is addressed in \texttt{GenericSolutionTruncation} and consists of all solutions starting with a non-critical curve point. 
Each critical curve point corresponds to a set (that could be empty) of particular Puiseux series solutions.\\
The command computes the generic solution, all particular solutions expanded at $0$ and all solutions expanded at infinity. 
The solutions are represented as truncations such that existence and uniqueness is ensured. 
In other words, the truncations are in one-to-one correspondence to the solutions. 
By setting the optional arguments genericsolution, const, computeFinite, computeInf to false, the corresponding subsets of the solution set can be suppressed. 
The remaining option iv$=y_0$ represents an initial condition of the format $y(0)=y_0$, where $y_0$ is an element of the ground field or an algebraic extension field of it, which is additionally taken into account.

\vspace*{2mm}

\begin{itemize}[noitemsep,topsep=0pt]
	\item[$\diamond$] Calling Sequence: \texttt{> SolutionTruncations(F, N, options)}
	\item[$\diamond$] Input: a polynomial $F$ in $y,y'$, a rational number $N$ (by default set to zero) and several optional arguments: genericsolution, const, computeFinite, computeInf (all boolean) and a constant iv.
	\item[$\diamond$] Output: a list consisting of three components: the generic solutions, the solutions expanded at $0$ and the solutions expanded at infinity represented as truncated Puiseux series (modulo $x^N$).
\end{itemize}

\para

\noindent
{\bf \texttt{> GenericSolutionTruncation}}\\
The first order differential equation  has a generic local solution $ y(x)=y_0+y_1 x+\mathcal{O}(x^2)$, where $F(y_0,y_1)=0$. 
If $F$ is irreducible as polynomial and $(y_0,y_1)$ is a regular affine point of the curve implicitly defined by $F$, the extension of $y_0+y_1x$ to a solution $y(x)$ is guaranteed and unique. 
The command \texttt{GenericSolutionTruncation} computes the first terms of the generic (formal) power series solutions, expanded around $0$, of the given differential equation.

Note that for every irreducible component one generic solution is computed. 
Thus, all generic solutions of $F=0$ are given by the union of the generic solutions of the components. 
If the given differential equation is known to be irreducible, the optional argument irreducible=true (see below) can be used in order to speed up computations.

The output of the command is a set of lists with two entries: a polynomial in $x$ representing the solution computed modulo $x^N$ involving an unspecified parameter $\_CC$ and a set of exceptional values for $\_CC$. 
For these values the generic solution would in general not lead to a solution of the given differential equation or might involve fractional exponents. 
Finally, let us mention that, if the precision of the output is not high enough, it is possible to use the command \texttt{ProlongSolutionTruncation}; see below.

\vspace*{2mm}

\begin{itemize}[noitemsep,topsep=0pt]
	\item[$\diamond$] Calling Sequence: \texttt{> GenericSolutionTruncation(F, N, options)}
	\item[$\diamond$] Input: a polynomial $F$ in $y,y'$, a rational number $N$ (by default set to zero), and optionally \texttt{irreducible} as boolean.
	\item[$\diamond$] Output: is a set of lists with two entries: a polynomial in $x$ representing the solution computed modulo $x^N$ involving an unspecified parameter $\_CC$ and a set of exceptional values for $\_CC$.
\end{itemize}

\para

\noindent
{\bf \texttt{> ProlongSolutionTruncation}}\\
For the given first order differential equation, if an appropriate change of variables $z(x)=y(x)+s(x)$ is performed, the resulting equation
$$G(x,z(x),z'(x))=F(y(x)+s(x),y'(x)+s'(x))=0$$
might be of Briot-Bouquet type. 
In case that $s(x)$ is such a solution truncation of $y(x)$, existence and uniqueness of the solution $z(x)$ of $G$ are ensured and the following coefficients can be found by a Newton type algorithm. 
In particular, this is the case when $s(x)$ is an output element of \texttt{GenericSolutionTruncation} or \texttt{SolutionsTruncations}.

In this situation, the command \texttt{ProlongSolutionTruncation} prolongs the first terms of a truncated Puiseux series solution $s(x)$ of $F(y(x),y'(x))=0$.

\vspace*{2mm}

\begin{itemize}[noitemsep,topsep=0pt]
	\item[$\diamond$] Calling Sequence: \texttt{> ProlongSolutionTruncation(F, s, N, x0)}
	\item[$\diamond$] Input: a polynomial $F$ in $y,y'$, a polynomial $s$, a rational number $N$ and $x0$ equals  $0$ or infinity (by default set to zero)
	\item[$\diamond$] Output: it is again a truncated Puiseux series computed until the order $x^N$ (or $1/x^N$)
\end{itemize}

\para

\noindent {\bf \texttt{> AlgebraicSolution}}\\
Algebraic solutions of the first order autonomous differential equation are represented by its minimal polynomisl, say $G(x,y)$. 
In this case, all the functions $y(x)$ with $G(x,y(x))=0$ are solutions of this differential equation and can be represented as Puiseux series.

Assuming that $F$ is an irreducible polynomial, the existence of algebraic solutions can be decided and, in the affirmative case, all solutions are algebraic and are given as shift of the independent variable, namely by $G(x+c,y)$. 
Therefore, by factorizing the given differential equation, all algebraic solutions can be found using this procedure for every component. 

The command \texttt{AlgebraicSolution} decides the existence of algebraic solutions of the given differential equation. 
Furthermore, if a solution exists the output is the minimal polynomial of the solution. 
The other solutions then can be easily found by shifting $x$. 
The solutions are found by checking whether a particular solution is algebraic. 
Efficiency of the algorithm highly depends on the chosen initial value. 
The procedure is using a formal power series solution, which means non-negative integer exponents for the solution, with a relatively small number of algebraic extensions of the ground field. 
Similarly to the command \texttt{GenericSolutionTruncation}, if the given differential equations is known to be irreducible, this can be specified by the optional argument irreducible=true.

\vspace*{2mm}

\begin{itemize}[noitemsep,topsep=0pt]
	\item[$\diamond$] Calling Sequence: \texttt{> AlgebraicSolution(F, options)}
	\item[$\diamond$] Input: $F$ is a first-order differential polynomial, and 
	\texttt{irreducible} is a boolean option.
	\item[$\diamond$] Output: the decision on the existence of algebraic solutions of  the differential equation. If a solution exists the output is the minimal polynomial of the solution.
\end{itemize}

\subsection{Usage of the Package}
In order to use the package, download the file FirstOrderSolve.m from\\ \texttt{https://risc.jku.at/sw/firstordersolve/} and save it as your local folder. 
After starting Maple you redefine the variable libname as

\texttt{> libname:=libname, 'path of user local folder';}

\noindent
Then, after executing the command

\texttt{> with(FirstOrderSolve);}

the package can be used. 
In the appendix we provide a Maple Worksheet illustrating the usage of our package for the computation of all Puiseux series solutions of first-order autonomous ordinary differential equations. 
We provide at \texttt{https://risc.jku.at/sw/firstordersolve/} an extended version of this file.

\appendix
\newgeometry{left=1mm,right=1mm,top=5mm,bottom=1mm}
\begin{figure}
\includegraphics[width=\textwidth,height=\textheight]{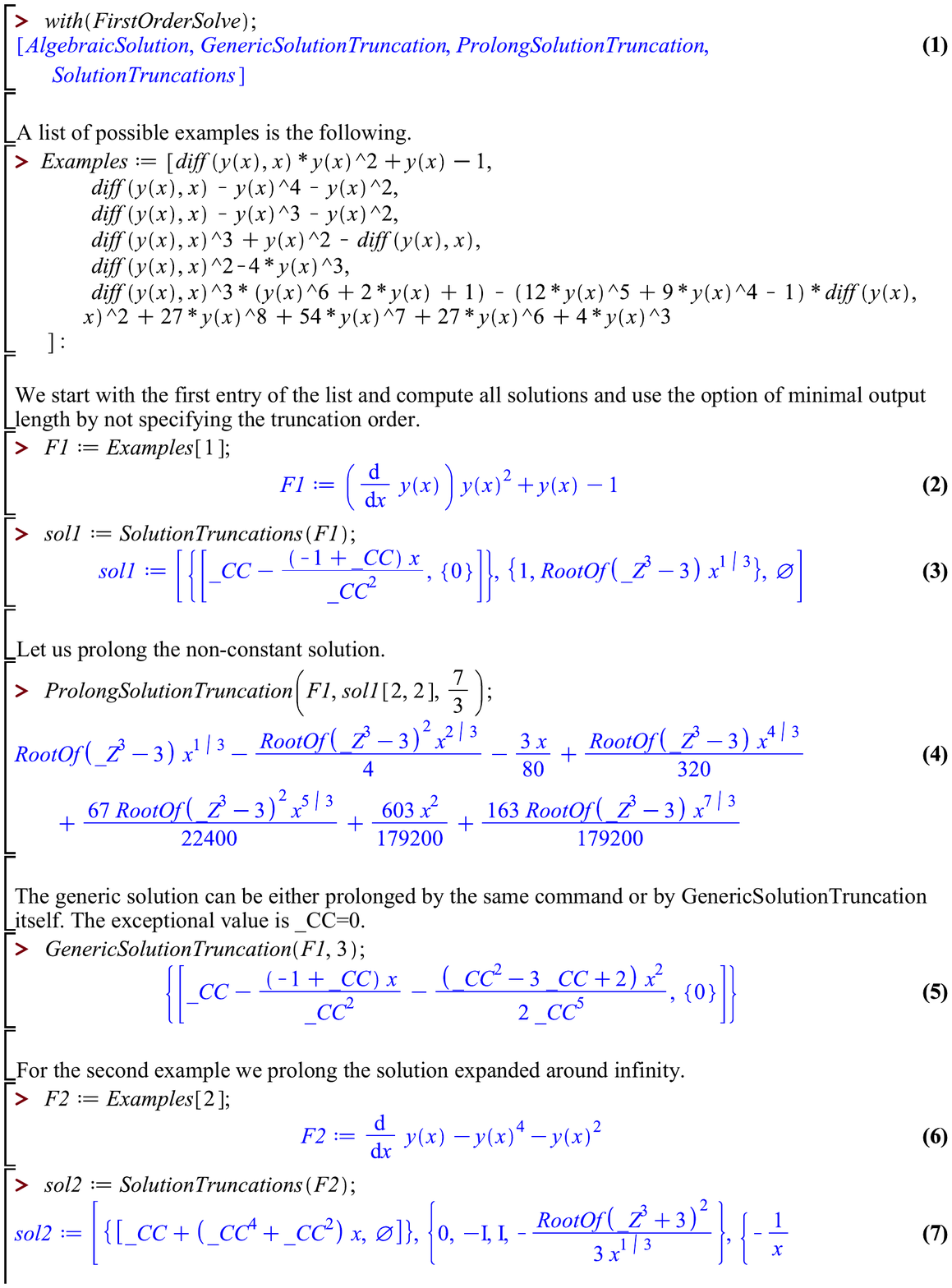}
\end{figure}
\begin{figure}
\includegraphics[width=\textwidth,height=\textheight]{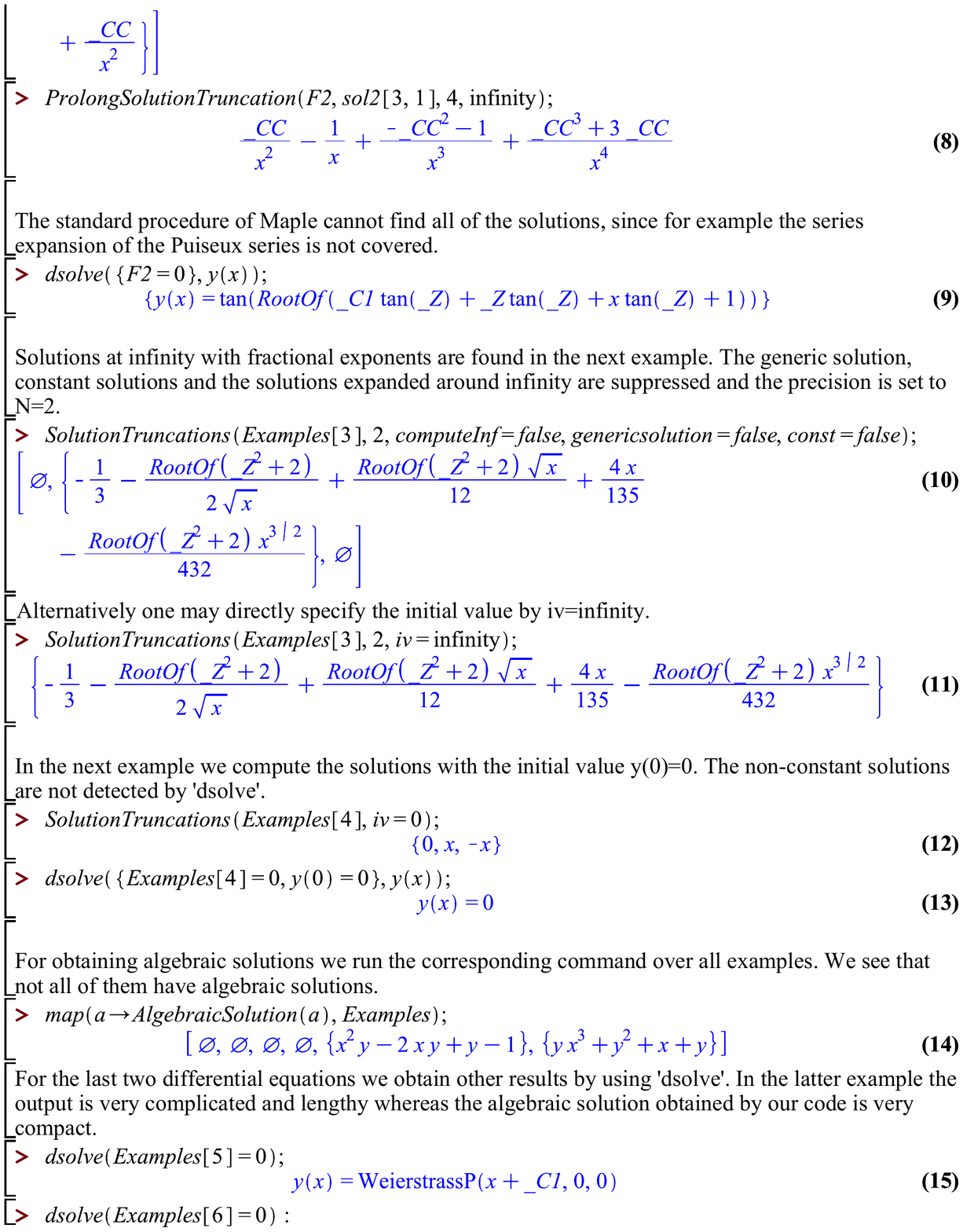}
\end{figure}
\restoregeometry

\bibliographystyle{alpha}

\begin{thebibliography}{}
	
	\bibitem[ACFG05]{aroca2005algebraic}
	J.M. Aroca, J.~Cano, R.~Feng, and X.-S. Gao.
	\newblock Algebraic {G}eneral {S}olutions of {A}lgebraic {O}rdinary
	{D}ifferential {E}quations.
	\newblock In {\em Proceedings of the 2005 international symposium on Symbolic
		and algebraic computation}, pages 29--36. ACM, 2005.
	
	\bibitem[Aro00]{Aroca2000}
	J.M. Aroca.
	\newblock {\em {Puiseux Solutions of Singular Differential Equations}}, pages
	129--145.
	\newblock Birkh{\"a}user Basel, Basel, 2000.
	
	\bibitem[BB56]{BriotBouquet-Reserches}
	C.A. Briot and J.C. Bouquet.
	\newblock Recherches sur les propriet{\'e}s des {\'e}quations
	diff{\'e}rentielles.
	\newblock {\em Journal de l'Ecole Polytechnique}, 21:36:133--198, 1856.
	
	\bibitem[Can05]{Cano2005}
	J.~Cano.
	\newblock {The Newton Polygon Method for Differential Equations}.
	\newblock In {\em Proceedings of the 6th International Conference on Computer
		Algebra and Geometric Algebra with Applications}, IWMM'04/GIAE'04, pages
	18--30, Berlin, Heidelberg, 2005. Springer-Verlag.
	
	\bibitem[CFS20]{cano2019existence}
	Jose Cano, Sebastian Falkensteiner, and J~Rafael Sendra.
	\newblock {Existence and Convergence of Puiseux Series Solutions for First
		Order Autonomous Differential Equations}.
	\newblock {\em Journal of Symbolic Computation}, 2020.
	
	\bibitem[DDRJ97]{DoraJung1997}
	J.~Della~Dora and F.~Richard-Jung.
	\newblock About the {N}ewton {A}lgorithm for non-linear {O}rdinary
	{D}ifferential {E}quations.
	\newblock In {\em Proceedings of the 1997 International Symposium on Symbolic
		and Algebraic Computation}, ISSAC '97, pages 298--304, New York, NY, USA,
	1997. ACM.
	
	\bibitem[Duv89]{Duval1989}
	D.~Duval.
	\newblock {Rational {P}uiseux {E}xpansion}.
	\newblock {\em Compositio Mathematica}, 70(2):119--154, 1989.
	
	\bibitem[Fal20]{FalkensteinerThesis}
	S.~Falkensteiner.
	\newblock {\em {Power Series Solutions of AODEs - Existence, Uniqueness,
			Convergence and Computation}}.
	\newblock PhD thesis, RISC Hagenberg, Johannes Kepler University Linz, 2020.
	
	\bibitem[FG04]{feng2004rational}
	R.~Feng and X.-S. Gao.
	\newblock Rational {G}eneral {S}olutions of {A}lgebraic {O}rdinary
	{D}ifferential {E}quations.
	\newblock In {\em Proceedings of the 2004 international symposium on Symbolic
		and algebraic computation}, pages 155--162. ACM, 2004.
	
	\bibitem[FG06]{feng2006polynomial}
	R.~Feng and X.-S. Gao.
	\newblock A polynomial time {A}lgorithm for finding rational general solutions
	of first order autonomous {O}{D}{E}s.
	\newblock {\em Journal of Symbolic Computation}, 41(7):739--762, 2006.
	
	\bibitem[Fin89]{fine1889}
	H.~Fine.
	\newblock On the {F}unctions defined by {D}ifferential {E}quations, with an
	{E}xtension of the {P}uiseux {P}olygon {C}onstruction to these {E}quations.
	\newblock {\em American Journal of Mathematics}, 11:317--328, 1889.
	
	\bibitem[Fin90]{fine1890}
	H.~Fine.
	\newblock Singular {S}olutions of {O}rdinary {D}ifferential {E}quations.
	\newblock {\em American Journal of Mathematics}, 12:295--322, 1890.
	
	\bibitem[GS91]{GrSi:1991}
	D.Y. Grigoriev and M.~Singer.
	\newblock Solving {O}rdinary {D}ifferential {E}quations in {T}erms of {S}eries
	with {R}eal {E}xponents.
	\newblock {\em Trans A.M.S.}, 327:329--351, 1991.
	
	\bibitem[Sta00]{RISC4119}
	P.~Stadelmeyer.
	\newblock {\em {On the {C}omputational {C}omplexity of {R}esolving {C}urve
			{S}ingularities and {R}elated {P}roblems}}.
	\newblock PhD thesis, RISC, Johannes Kepler University Linz, 2000.
	
	\bibitem[VGW18]{RISC5589}
	N.T. Vo, G.~Grasegger, and F.~Winkler.
	\newblock {Deciding the Existence of Rational General Solutions for First-Order
		Algebraic ODEs}.
	\newblock {\em Journal of Symbolic Computation}, 87:127--139, 2018.
	
	\bibitem[Wal50]{walker1950algebraic}
	R.J. Walker.
	\newblock {\em Algebraic {C}urves}.
	\newblock Princeton University Press, 1950.
	
\end{thebibliography}

\end{document}